\documentclass[sigconf]{acmart}

\renewcommand\footnotetextcopyrightpermission[1]{} 
\makeatletter
\renewcommand\@formatdoi[1]{\ignorespaces}
\makeatother
\acmConference[AI4FashionSC'20]{The fifth international workshop on fashion and KDD}{August 23, 2020}
{Virtual Event, CA, USA}

\AtBeginDocument{%
  \providecommand\BibTeX{{%
    \normalfont B\kern-0.5em{\scshape i\kern-0.25em b}\kern-0.8em\TeX}}}

\begin{document}

\title{Explainable AI based Interventions for Pre-season Decision Making in Fashion Retail}

\author{Shravan Sajja}
\email{suryasku@in.ibm.com}
\affiliation{%
  \institution{IBM Research India}
}

\author{Nupur Aggarwal}
\email{nupaggar@in.ibm.com}
\affiliation{%
  \institution{IBM Research India}
}

\author{Sumanta Mukherjee}
\email{sumanm03@in.ibm.com}
\affiliation{%
  \institution{IBM Research India}
}
\author{Kushagra Manglik}
\email{kmangli1@in.ibm.com}
\affiliation{%
  \institution{IBM Research India}
}

\author{Satyam Dwivedi}
\email{satydw10@in.ibm.com}
\affiliation{%
  \institution{IBM Research India}
}

\author{Vikas Raykar}
\email{viraykar@in.ibm.com}
\affiliation{%
  \institution{IBM Research India}
}

\renewcommand{\shortauthors}{Sajja, et al.}

\begin{abstract}
Future of sustainable fashion lies in adoption of AI for a better understanding of consumer shopping behaviour and using this understanding to further optimize product design, development and  sourcing to finally reduce the probability of overproducing inventory. Explainability and interpretability  are highly effective in increasing the adoption of AI based tools in creative domains like fashion. In a fashion house, stakeholders like buyers, merchandisers and financial planners have a more quantitative approach towards decision making with primary goals of high sales and reduced dead inventory. Whereas, designers have a more intuitive approach based on observing market trends, social media and runways shows. Our goal is to build an explainable new product forecasting tool with capabilities of interventional analysis such that all the stakeholders (with competing goals) can participate in collaborative decision making process of new product design, development and launch. 
\end{abstract}


\maketitle

\section{Introduction}

One of the biggest problems confounding a lot of fashion houses is the problem of unsold dead inventory. Despite having a lot of historical sales data, most fashion houses are able to sell only 50-60\% of products/Stock Keeping Units (SKUs)  well and the rest go through severe mark downs. Since fashion is heavily trend driven and most retailers operate by season (for example, spring/summer, autumn/winter, holiday etc.), at the end of each season any unsold inventory is generally liquidated. While smaller retailers generally move the merchandise to second-hand shops or e-commerce companies; large brands resort to recycling or destroying the merchandise.

Unsold merchandise/inventory is mainly due to mis-match between supply and demand in the supply chain network. It could be that the inventory has been over-produced or the inventory has not been distributed properly at the right location and at the right time, mainly due to inaccurate demand forecasts. At the heart of the problem we need to be able to accurately forecast the sales or sell-through-rate (STR) of any existing product/SKU or a potential new product at any given {\it location} and {\it time}. Unlike other retail industries fashion is heavily trend driven and in every season a substantial amount of new products are introduced.  There is no reference historical sales data for these new products to base the sales prediction on. For example, consider a retailer who wants to forecast sales for never-out-of-stock (NOOS) product like {\it white shirts}, for the next month. Typically the retailer has access to the historical sales of {\it white shirts} for the past few years. Standard time-series forecasting methods can be applied to forecast the sales(demand) based on the historical time-series data. However, consider a fashion retailer who is introducing a new {\it botanical print sleeveless top} and wants to forecast the demand for this product to better manage the supply chain. Since this is a completely new product there is no explicit historical sales data to base the forecast on. Hence, there is inherent risk associated with new product forecasting and introducing new products in the market. Notwithstanding the risk associated with new products, fashion houses typically introduce 60-80\% new products every season as a business strategy to satisfy consumer demand at its peak. It enables the fashion house to stay relevant and preserve their customer base in a world of constantly changing fashion trends and demands. 

Despite, all the developments in the area of AI enabled forecasting methodologies, most stakeholders in the fashion supply chain (designers, buyers, merchandisers) still resort to a gut based approach of averaging sales of similar products for new product forecasting. This approach is more popular because of its interpretability. Sales forecasting is a company-wide process used by multiple stakeholders to guide many operational decisions and interventions.  Thus, it is desirable to build forecasting models which are  trustworthy, interpretable and transparent. The primary goal of this project is to build an {\it explainable} sales analysis and forecasting asset that involves analyzing past sales and forecast sales for new products at different stages of product design and development. Explainable sales forecasting models will not only improve the trustworthiness of model's outputs but would also provide transparency for all stakeholders involved in the process of new product development and launch. Thus, improving accountability and fostering a collaborative environment amongst stakeholders with competing requirements. We use multiple tools of global, local explainability and product life cycle management to enhance interaction between the forecasting asset and the stakeholders.  The second goal of this project is to enable product developers and designers to perform pre-season interventions on product design. These pre-season interventions are based on tools like what-if analysis and counterfactual explanations; and these interventions are surfaced through an intuitive user-interface. Pre-season interventions drive the product through multiple iterations/versions and a new product forecast is generated for every version of the product until a product with high potential STR is achieved. In the next section we provide a brief background on product design and development, planning and buying. We also introduce several stake holders and their roles in the fashion supplychain. Section \ref{sec:related_work} describes related work in the areas of new product forecasting and explainable machine learning methods. In section \ref{sec:exp_sales_analysis} we present different aspects of our explainable sales analysis module and in section \ref{sec:use_cases} we describe the use cases specific to different stakeholders. Finally, in section \ref{sec:conclusion} we reflect on our experiences in building this asset and using it in conjunction with our industry partners.



\section{Fashion personas, product design and development}\label{sec:background}

Empirical data for this paper is obtained from our industry partner which happens to be a fashion house in India  that targets women in the 16-24 years age group. For this house, product development starts with the creation of an option plan for a  target season with details of  different categories (like t-shirts,tops,jeans,dresses etc.) and the number of unique styles (breadth) that needs to be developed for this season along with the number of units that need to be bought for each style (depth).  Option plans are developed by buyers and financial planners of the fashion house based on the historical sales data of their products and their future strategy. These decisions are further segregated for different price points (low, medium and high).  For this fashion house, each fashion season would start with an allocation of  80\% of their budget to cover the following three types of products:

\noindent {\bf 1. Commercial fashion} (75\%) - The main set of fashion products that interpret the current trends for the target customer. They are primarily new products, however, they do not deviate drastically from past products sold by the fashion house.
 
\noindent {\bf 2. Trend led high fashion} (20\%) - A heavily trend led product  often introduced for fashion credibility of the brand/label.
 
\noindent{\bf 3. Never-Out-Of-Stock (NOOS)} (5\%) - This department includes the core basics which are carried every season. There is no major product design and development for these type of products and fashion houses are primarily concerned with store level stock allocation and replenishment tasks for them.

 Around 20\% of the budget is set aside for the {\it express department} meant for in-season drops of styles which were extremely popular and have sold out, thus requiring a re-order from the vendors. There exists another category of merchandise called Old Season Merchandize (OSM), which as the name suggests are products from previous season and they have been continued in the stores they exceptionally well and there is continued demand for it to be carried into the next season or
 the salvage price of a particular product is prohibitively low. In this paper we focus on forecasting performance of products belonging to the commercial fashion category. Preparation of option plan and budget planning is beyond the scope of this paper.

Designers for this fashion label work on roughly 12 collections in a year and hence they follow a monthly calendar. They begin their product development and design process with a trend study for the target season 6 months prior to delivery of products to the store. Trend study is based on a global trend report from the parent company and other local trend reports curated by the designers. Local trend reports are more informal and they are typically based on designer's vendor visits, competitors, travel and observation of social media channels. More formal trend forecasts from WGSN and Stylumia are also incorporated in these trend reports. Designers interpret the trend reports design and develop a certain number of products for each category as specified in the option plan. Each initial product design is created as a {\it docket} which has a few reference product images, some attributes (like Pantone color reference, fabric, pattern, shape etc.), the desired vendor and any changes that needs to be communicated to the vendor. Based on the required depth in the option plan and the minimum order quantity (MOQ) the designer initially decides on the right vendor for the product. The docket can also include the measurement chart or a physical sample. The docket is then sent to the vendor who then creates a physical {\it sample} and sends it back to the designer. It typically takes around 3 weeks to procure a sample from the vendor. Based on the samples, the designer curates and presents the designs as a range board to the buyers. The range board is organized around collections and stories and there are also indications of which designs should be bought together for the collection to make sense.

The buyer curates the designs presented by the designer based on their estimation of how well the product will sell (based on historical sales data and his/her interpretation of trends). The buyer can also go back to the designer for any changes, reject designs or asks for new designs. Eventually he/she creates a final order sheet specifying products to be bought for this season. At this stage a final {\it techpack} is created for each product which is an excel sheet/document for each product with the style images, detailed product attributes, the desired vendor and any changes that needs to be communicated to the vendor. The {\it techpack} essentially has all details for production. Store-level new product forecasting  will be further applied to {\it techpacks} to make merchandising decisions about hyper-local assortment and stock allocation of products. Hyper-local assortment and stock allocation are critical pre-season decisions for a fashion house, however, considering space constraints, we do not focus on these aspects in this paper.

This project primarily focuses on improving the above mentioned process involving designers and buyers. They are important stakeholders and decision makers in the product life cycle and both personas use new product forecasting to guide their decisions. New product forecasting begins at the {\it docket} stage and its predictions, uncertainty, explanations and counterfactual explanations will be used by designers to make appropriate changes to possibly improve the product and move towards the {\it sample} stage.  At the {\it sample} stage, both buyers and designers use new product forecasting during the range show. A common explainable forecasting platform for the {\it sample} will enable buyers and designers to collaboratively prepare the final product range. An important metric to measure the performance of our system is the hit -rate achieved by designers. {\it Hit-rate is the measure of efficiency for designers in terms of number of designs accepted by the buyer against the total number of designs produced.} Higher hit-rates are indicative of higher synergy between designers and buyers.

\vspace{-3mm}
\section{Related work}\label{sec:related_work}
The problem of new product forecasting has received increased attention in recent years and the papers \cite{liu2013sales, chase2013demand, baardman2017leveraging} provide a good survey of different new product forecasting methods. Even though, there is little to no data for new products, note that there is historical sales data for set of other products from last season or last week. In literature, most machine learning based methods for new product forecasting, focus on proxies such as demand/sales of similar products in the past. For example, given a new {\it botanical print sleeveless top}, the merchandiser would identify similar products in the last season and forecast the sales of a new product as an average of the sales of all those similar products.  The design team would further observe social media, fashion ramps, competing fashion labels  and manually adjust demand forecast for this particular {\it botanical print sleeveless top}. It was reported in \cite{garro2011new} that ZARA would follow a similar approach, where, similar products were chosen from the current season by buyers, distribution staff and store managers. Similar products were chosen from the current season to capture the latest trends. This approach was necessary and possible because ZARA would drop new products every fortnight, hence data from the previous weeks would inform the forecast for similar products this week. 

Most forecasting methods for {\it evolutionary} new products follow a similar approach. Evolutionary new products are characterized by incremental improvements and evolutionary changes in the product line, nevertheless majority of new product launches in the fashion industry are evolutionary. Hence, the forecasting methods used for new fashion products, primarily differ in how similar products (or proxies) are selected. In \cite{thomassey2006hybrid}, similar products are obtained by clustering existing products based on their historical sales curves and all the products in a cluster share a forecasting model. Authors in \cite{baardman2017leveraging} propose a cluster-while-regress approach  where clusters are formed based on  similarity in terms of both product features as well as sales behavior of the products. Contrary to the belief that similar products produce similar performances, authors in \cite{singh2019fashion} observe that, products with same date of launch performed similarly. Thus, relative placement of products and its combination with product features plays a significant  part in product's performance. Hence, it is necessary to build a model which is able to identify similar behaviour as a nonlinear function of product's input features.







\vspace{-2mm}
\section{Methodology}

In this section, we introduce some of the challenges we faced during this project and how these challenges guided our decisions and methodology. Sales data is affected by multiple factors and machine learning models are capable of using product design attributes (color, fabric, pattern, fit, size, neckline), merchandising factors (markdowns, promotions, visibility), store-locations, weather, holidays, events, market-sentiment for similar products, prices of competing products within the store and in the market and macro-economic factors. Since future sales cannot be modeled purely based on sales history alone, we include most of the above mentioned features in our modeling process. Effect of location is encoded in the sales data at the store level. Hence, for this paper we do not focus on weather and demographics of a location explicitly.  

\noindent {\bf 1. Forecasts should be explainable:} Forecasting sales/demand accurately for new products is a challenging task with track record of very high error percentages \cite{chase2013demand} and the merchandisers typically have a strong gut feeling of the industry trends based on their past experience. Hence unless our model is able to explain the forecast merchandisers will be very reluctant to trust the forecast from an algorithm.  Hence, our primary requirement in this project is explainability: we would like to explain both in terms of the intrinsic attributes and also the exogenous regressors. Both global (What are the factors affecting the overall demand?) and local explainability (Why was the demand high for this particular product?) are important. For exogenous regressors explanations should be contrastive with reference to the baseline forecast without any external events.

\noindent {\bf 2. Structured data and missing values:} Product features are recorded faithfully during the design process, hence their combined effect on sales of individual products can be modelled using regression models. Any irregularity in capturing design attributes from product meta data can be handled through standardization using our fashion taxonomy. However, standardization of attributes over a category of fashion apparel may lead to {\it missing} attribute values for particular products. Hence, our approach should be able to handle missing data during the training as well as prediction tasks. In case structured attributes are not available, then it could as well be neural network based embedding for the product images or textual descriptions, however, image based new product forecasting is beyond the scope of this project.

\noindent {\bf 3. Dis-aggregation of merchandising factors:} Based on the data we received from our industry partner, the effect of merchandising features like markdowns, volume discounts and promotions on sales of individual products is more difficult to quantify. Primary reason being, the problem of segregating volume discounts and markdown on individual products from point-of-sale (POS) data. When consumers buy multiple products at the same time, they receive volume discounts on their baskets and the values of markdowns on individual products are not recorded in the POS data. Hence, inclusion of markdown in the modeling process may result in spurious relationships. 


\noindent {\bf 4. Holidays, events, demand transference:} Public holidays, festivals and special events have direct impact on performance of a product. Hence, time of launch of a product becomes a critical factor in deciding the performance of a product in the market. Using the week of launch as a product attribute will capture the effect of holidays and festivals. Using the week of launch as a product attribute will also allow us to capture the presence of all the other products that were available in the store in that week. Thus, allowing us to implicitly model interaction between these products, more formally known as demand transference.

\noindent {\bf 5. Sell-through-rate versus sales:} Sales alone doesn't represent performance of a product because sales data doesn't include information about store-level inventory, stock-outs and inter-store transfers.  Sell-Through-Rate (STR) measures the amount of inventory sold versus the amount of inventory received from the supplier.  STR values of individual products are monitored closely to enable {\it in-season} interventions like markdowns, inter-store transfers and withdrawal of the product to a central distribution center (DC). However, superficial reading of STR values can result in erroneous judgements. For example, towards the end of a season, pull-back of a product due to bad performance will result in low inventory values and hence higher STR values. So higher STR values for products towards the end of season do not represent high product performance. In this paper, we consider the STR value of a product at the end of first 4 weeks from its launch date as a more legitimate indicator of its performance and as the target variable of our model. Because, our industry partners start marking down products after 4 weeks of its launch. So STR values after the $4^{th}$ week will be corrupted by markdowns, inter-store transfers and product withdrawals.  \newline

\noindent The main assumption we make is that each product $p$ can be represented as a vector of $d$ features/attributes $\mathbf{x}_p \in \mathbb{R}^d$. Design attributes like color, pattern, sleeve style etc. are transformed into numerical variables using label transformers or numeric encoders. Merchandising attributes like list price of the product is included in $\mathbf{x}$. Associated with each product we have a historical time series for target variable (sales/demand/STR):
$$\mathbf{y}_p = [y_p(t_{p0}), y_p(t_{p0}+1), y_p(t_{p0}+2), \cdots, y_p(t_{p0}+T_p)]$$
\noindent where the interval $[t_{p0}, t_{p0}+T_p]$ represented the entire life cycle of  product $p$ in a season. The time instant $t=t_{p0}$ represents the product launch date and $T_p$ represents the date untill when that product is carried in the stores. The interval $[t_{p0}, t_{p0}+1]$ can be chosen based on the nature of product being analyzed, for this paper we operate on weekly basis. Thus, $y_p(t_{p0})$ would represent the aggregate value of our target variable over the first week of product's launch. Given a set of $n$ products, we have
$$\mathcal{D}=\left[(\mathbf{x}_1,\mathbf{y}_1),(\mathbf{x}_2,\mathbf{y}_2),...,(\mathbf{x}_n,\mathbf{y}_n)\right]$$ 
the task is to learn a model $f$ to predict the demand $\mathbf{y}_{\sf new}(T)$  for a new product $\mathbf{x}_{\sf new}$ given by 
$\mathbf{y}_{\sf new}=f(\mathbf{x}_{\sf new}\,\vline \, \mathcal{D})$. In this paper, we consider first 4 week sell through rate as our target variable, hence, $\mathbf{y}_p =  STR_p(t_{p0}+3)$
and $\mathcal{D}$ is given by
$$\left[(\mathbf{x}_1,STR_1(t_{10}+3)),(\mathbf{x}_2,STR_2(t_{20}+3)),...,(\mathbf{x}_n,STR_n(t_{n0}+3))\right].$$

\noindent This modelling framework is valid for all levels of hierarchy like store, city, region, country. As we move up the hierarchy, target variables get spatially aggregated. A store-level model will capture the effect of location, i.e., implicitly capture the effect of weather and demographics on the target variable. \newline

\noindent The effect of {\it temporal variables} like time of product launch, holidays, events etc. on the target variable is captured by appending the time of launch $t_{p0}$ (launch week or launch month)  to the product attribute vector $\mathbf{x}_p$  given by $\mathbf{z}_p = \begin{bmatrix}\mathbf{x}_p \quad t_{p0}\end{bmatrix}^T$. Since, our target variable is STR in the first 4 weeks of a product launch, {\it interventional variables}  like discount and promotions need not be included in the modelling framework. Finally, we have $\mathcal{D}_z =$ 
$$\left[(\mathbf{z}_1,STR_1(t_{10}+3)),(\mathbf{z}_2,STR_2(t_{20}+3)),...,(\mathbf{z}_n,STR_n(t_{n0}+3))\right].$$ 
Now the task is to learn a model $f_z$ to predict $STR_{\sf new}(t_{\sf new}+3)$  for a new product $\mathbf{x}_{\sf new}$, i.e., $
\mathbf{y}_{\sf new}=f_z\left([\mathbf{x}_{\sf new},t_{\sf new}]^T \,\vline \,\mathcal{D}_z\right)$ where  $t_{\sf new}$ is the new product launch week.

\noindent {\bf Fashion taxonomy:}  Consider a new product with id {\bf XYZ} under the category {\bf shirts}, then we have a following hierarchy:
$$\bf all \, categories \longrightarrow category: \, shirts \longrightarrow  shirts\, similar \, to \, XYZ$$
\noindent This modelling framework is valid for all levels of such hierarchies. We also studied product similarity based on visual search and natural language search.
\begin{itemize}

\item {\it Visual Browse/Search}: For any given image of a new product we can find other similar images in the product catalog purely based on visual appearance \cite{jing2015visual}.

\item {\it Natural Language Search (NLS)}: The Natural Language Semantic Search  allows the user to search a fashion e-commerce catalog using complex natural language semantic queries. For example, our system supports natural language queries like "red tops", "floral dresses" etc. 
\end{itemize}

\noindent Results of these searches can be used to collect all the similar items  for a new product. However, building an explainable model at the level of {\bf shirts similar to XYZ} was challenging because of limited data points. Hence, for this paper we focus on building new product sales forecasting models at the {\bf category} level.
\vspace{-3mm}
\subsection{Predictive model}

\noindent eXtreme Gradient Boosting or XGBoost is an optimized distributed gradient boosting library designed to be highly efficient, flexible and portable \cite{chen2016xgboost}. It  has achieved state-of-the-art results in many areas due to its scalability and it is known to have advantages over other tree based learners both in terms of speed and prediction accuracy \cite{al2019comparison}.  Based on the assumptions, challenges and requirements presented earlier, like structured data, explainability, ability to handle missing values and ability to handle nonlinear relationships we chose the two tree based learners: XGBoost and CatBoost \cite{prokhorenkova2018catboost}, to learn $f_z$. CatBoost is also an open-source machine learning algorithm with categorical features support and it is known to yield state-of-the-art results without extensive data training. A good comparison of these methods can be found in \cite{al2019comparison}. Another important feature of XGBoost and CatBoost methods is their instance weighting capability. It is necessary for us in scenarios where there is an imbalance in distribution of high selling and low selling products, because both these categories of products can share similar sell through rates. For example, sales of 4 units given 5 units of inventory will produce the same STR for the scenario when 400 units were sold with 500 units of inventory in the store. This issue can be addressed by weighing each data instance of the training data by normalized values of sales for each product. Both XGBoost and CatBoost provide a weight parameter during the training phase to satisfy this requirement.


\vspace{-3mm}
\subsection{Distribution estimator}  
Each point forecast  $y = f(\mathbf{x})$ should be associated with a measure of certainty for that value. Most stochastic optimization algorithms in supplychain use-cases, benefit from having a prediction interval to quantify the uncertainity in the prediction. Thus, generating prediction intervals for predicted variable is a common practice in machine learning domain. These prediction intervals are generated through quantile regression, however, this approach requires us to generate a model for each quantile that a user is interested in. User requirements vary and a demand for multiple quantiles can arise, in such a case, training and managing all the models corresponding to all the quantiles may become untenable. Thus we take the approach of building a distribution estimator which models the errors of a prediction model and this estimator can generate prediction intervals for any given probability without re-training or recalibration.\newline

This approach assumes that $y|\mathbf{x}$ is a normal distribution. The base model) predicts the mean of the gaussian distribution, whereas the estimated error give us the standard deviation of the distribution. The data $(\mathbf{X},\mathbf{y})$ is split into $(\mathbf{Xb},\mathbf{yb})$ to train the base model and $(\mathbf{Xe},\mathbf{ye})$ to train the error model. A base model $f_b$ is first trained on the $(\mathbf{Xb},\mathbf{yb})$. The squared prediction error $(\mathbf{ye}-f_b(\mathbf{Xe}))^2$ is computed on the validation set $(\mathbf{Xe},\mathbf{ye})$. The error model $f_e$ is then trained on $(\mathbf{Xe},(\mathbf{ye}-f_b(\mathbf{Xe}))^2)$ to regress on the squared error. For any new instance $\mathbf{x}$ the mean prediction is given by $f_b(\mathbf{x})$ and the 90\% prediction interval is given by $[f_b(\mathbf{x})-1.64\sqrt{f_e(\mathbf{x})},f_b(\mathbf{x})+1.64\sqrt{f_e(\mathbf{x})}]$. Thus, distribution estimator needs only 2 models.

\vspace{-3mm}
\subsection{Data}
A retail supply chain typically consists of 4 main databases:

\begin{itemize}
    \item {\it Sales}: This refers to the Point-of-sales (POS) transaction data from stores or other channels. Each document in the database captures a transaction of the following form: Customer A bought B units of product C at store/channel D for price E at date T.
    \item {\it Inventory}: This refers to the inventory data from stores or other channels for different products. Each document in the database captures a {\it inventory} snapshot of the following form: Inventory for product C at store/channel D at date T was X.
    \item {\it Stores}: This refers to the certain store level details like store address, latitude and longitude, store capacity, floor layouts, selling square footage etc.
    \item {\it Products}: This refers to the product meta-data (category,attributes and description) and corresponding product images.
       \end{itemize}
There are other optional databases like customer databases,  festival calendars and {\it exogenous variables}  like weather  etc. which are useful for sales forecasting. We also need a record of all the {\it interventional variables} like markdowns on individual products, volume discounts, discounts on super Saturdays, black Friday sales, Boxing day sales, discounts on mid season and end of season sales, etc. Most modern retail supply chains are highly integrated, hence, the process of extracting, transforming and loading these multiple databases into our database is possible with moderate effort. However, these databases are seldom integrated and they need extra steps of manual data collection, pre-processing, validation for maintaining data hygiene.

\vspace{-3mm}
\subsection{Testing and evaluation}
XGBoost and CatBoost are popular machine learning models and they have been used to achieve state-of-the-art results in several domains. Both models satisfy all our constraints and requirements, hence, we only compare these two models for their accuracy on our data-sets. We train both models on data from Spring-Summer collection of 2019 from our industry partner and initially we consider the following categories: {\bf shorts} (100 products), {\bf pants} (346 products),  {\bf dresses} (445 products), {\bf jeans} (461 products), {\bf shirts} (483 products), {\bf knits} (516 products), {\bf tops} (1031 products) and  {\bf t-shirts} (1248 products). These models have been built by tuning hyper-parameters of  XGBoost and CatBoost using \texttt{GridSearchCV}  from Scikit-learn \cite{pedregosa2011scikit}.  XGBoost works only with numerical variables and hence it requires conversion of categorical variables into numerical variables. For this purpose, we compared multiple approaches: label encoder from Scikit-learn,  one-hot encoding and a numeric encoder \footnote{A numerical encoder converts the categorical variables to numbers from 1 to $K$, where $K$ is the number of unique classes of the categorical variable.}. The numerical encoder performed the best amongst all three approaches. \newline 

\begin{figure}[h]
  \centering
  
  \includegraphics[width=0.8\linewidth, height = 4cm]{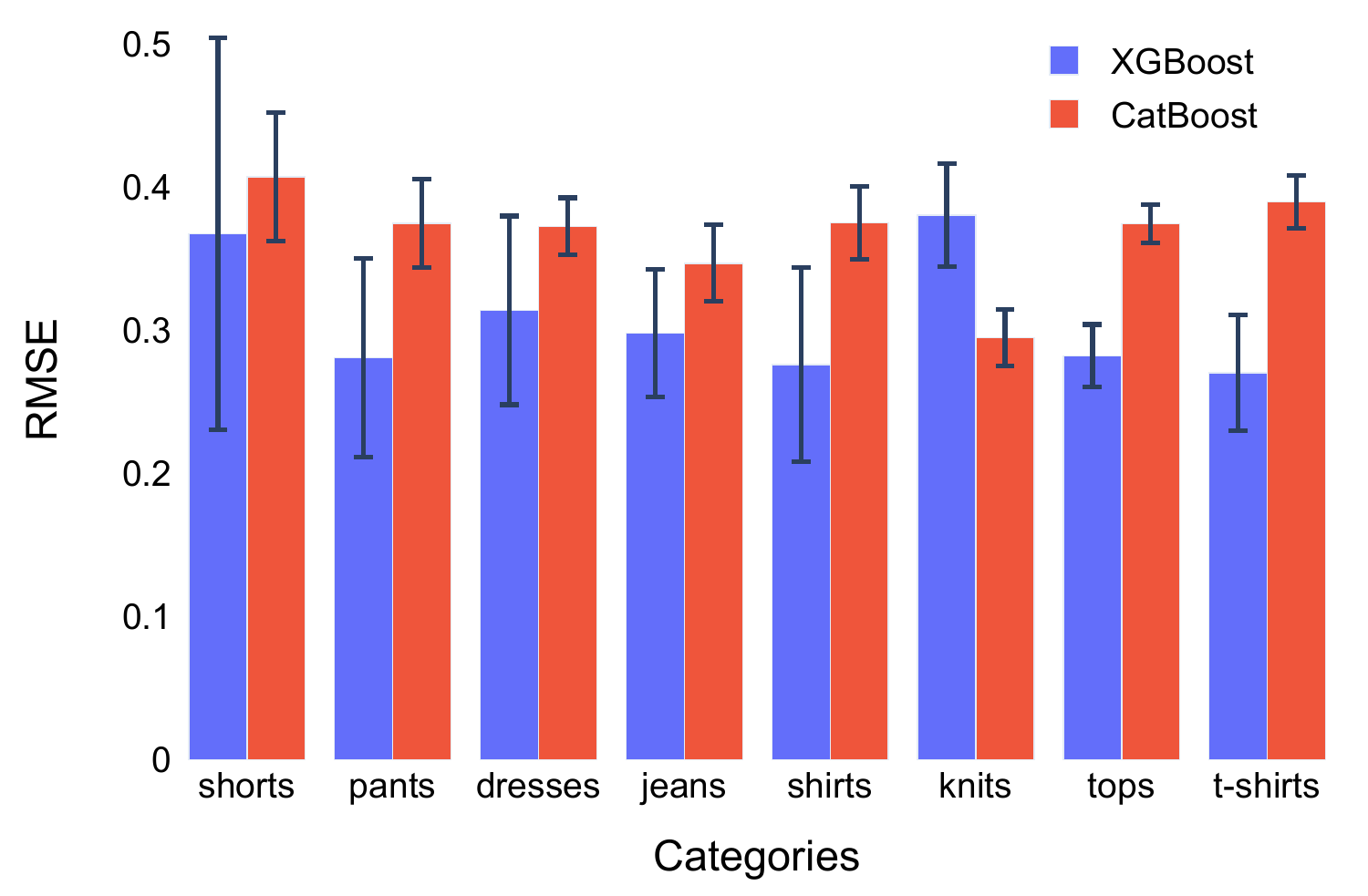}
  \vspace{-5mm}
  \caption{Comparison of XGBoost and  CatBoost models for different categories.}
\label{xgvscat}
      \vspace{-5mm}
\end{figure}

We compared these models for all categories on a test data-set chosen randomly from the original data-set. Root-mean-squared-error (RMSE) is used as our metric for error measurement. These results are presented in figure \ref{xgvscat}. It can be observed that, for category level models, XGBoost mostly performs better than CatBoost for our data-set and the selection of parameters. XGBoost starts performing better with increasing number of data points (see the categories: tops and t-shirts). Hence, we restricted ourselves to XGBoost models for this project.

\vspace{-2mm}
\section{Explainable sales analysis }\label{sec:exp_sales_analysis}

Explainability/Interpretability is the degree to which a human can understand the cause of a decision (or prediction) made by a prediction model. Our primary goal is to improve explainability of designers/buyers and allow them to perform interventions with forecasting-models in an interactive fashion.  There are two notions of explainability which we will typically use.

\begin{enumerate}
    \item {\bf Global Explainability for a model} - How do parts of the model (features/attributes) affect the predictions?

\item {\bf Local Explainability for a single prediction} - Why did the model make a certain prediction for an instance?
\end{enumerate}
While there are numerous notions of explainability available in the literature the core prediction algorithms used in our project support the following notions of explainability. Since our core prediction models are tree based (XGBoost, CatBoost) the discussion below is centered around explainability for tree based methods.
\vspace{-3mm}
\subsection{Factors affecting overall sales: Global explainability} \label{subsection:global} 
Global explainability of  a model provides insights into the sales potential of various categories and their corresponding design and merchandising attributes.

\subsubsection{Feature importance} 
\begin{figure}
  \centering
  \includegraphics[width=0.7\linewidth, height = 5cm]{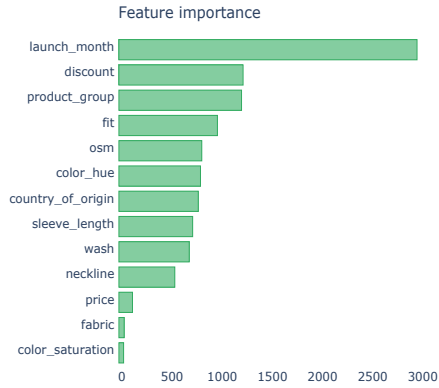}
      \vspace{-2mm}
  \caption{Feature importance plot generated using model based approach for tops}
    \vspace{-3mm}
    \label{fig:feature_imp_model}
\end{figure}

It is one of the  most widely used and
most well-studied explainability technique in literature \cite{gilpin2018explaining}. Feature importance provides the relative contribution of each feature to the model. We can derive feature importance values in multiple ways, however, in this paper we present two approaches:

\noindent \textbf{Model based approach}: Feature importance provides a score that indicates how useful or valuable each feature was in the construction of the boosted decision trees within the model. The more an attribute is used to make key decisions with decision trees, the higher its relative importance. Feature importance of XGBoost models can be derived from the importance matrix associated with the model. It consists of columns named: {\it weight}, {\it cover}, and {\it gain} and all these columns provide different global feature importance methods. However,  {\it gain} can be assumed to be the most relevant attribute to interpret the relative importance of each feature. 
The primary reason behind this assumption is due to the fact that {\it gain} column provides the relative contribution of a feature to the model and it is calculated by averaging over  contributions from all features for each tree in the model \cite{chen2016xgboost}. We plot the {\it gain } based feature importance scores for the category {\bf tops} in figure \ref{fig:feature_imp_model}. Feature names used in the figures are self explanatory and do not require description.

\begin{figure}[h]
  \centering
  \includegraphics[width=0.8\linewidth]{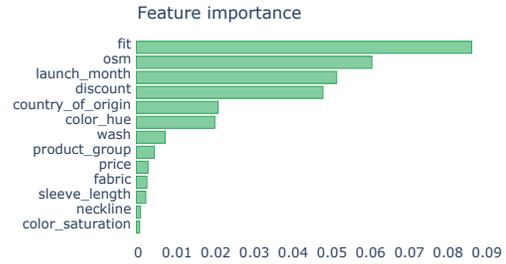}
      \vspace{-20mm}
  \caption{SHAP based feature importance plot for tops}
      \label{fig:feature_imp_shap}
    \vspace{-5mm}
\end{figure}

\noindent \textbf{SHAP based approach}: SHAP (SHapley Additive exPlanations) is a unified approach to explain the output of any machine learning model. SHAP connects game theory with local explanations \cite{lundberg2017unified}. SHAP explanations can also be used for global explanations to understand which features are most important for a model. For this we can compute the SHAP values of every feature for every sample. We can then take the mean absolute value of the SHAP values for each feature to get a feature importance score for each feature. Authors in \cite{lundberg2018consistent} evaluate different feature importance measures on the basis of their consistency and accuracy and they show that feature importance scores based on {\it gain} are inconsistent, i.e., there is no guarantee that feature with highest score is actually the most important.  Finally, they show that feature importance scores based on SHAP values are both consistent and accurate. We plot the SHAP value based feature importance scores for the category {\bf tops} in figure \ref{fig:feature_imp_shap}. Feature importance plots shown in figures \ref{fig:feature_imp_model} and \ref{fig:feature_imp_shap} differ in their feature attribution for the same model. To avoid confusion for designers/buyers we only present the SHAP values as our default feature importance plot. However, we provide the {\it gain} based plot option for more advanced users.

\subsubsection{Partial dependence plot} Forecasts should be interactive. We need to be cognizant of the fact that the buyer/merchandiser is eventually going to adjust the forecast given by the algorithm based on his/her gut instincts. In order to enable this collaborative forecast, the model should allow the user to play around with the attributes and exogenous regressors to asses the impact of them. For, example, What would be the change in the forecast if I introduce a markdown on a Wednesday? For example, a designer would be interested in knowing what would be the change in the forecast if he/she replaces the round-neck of a product with V-neck. This is possible through what-if analysis. What-if-analysis on product attributes will enable designers and buyers make informed decisions for their product attributes. The partial dependence plot (PDP) for a feature shows the marginal effect the feature has on the predicted outcome of a machine learning model.

\noindent \textbf{Model based approach}: The partial dependence plot (PDP) shows how the average prediction in your data-set changes when a particular feature is changed. The partial dependence function at a particular feature value represents the average prediction if we force all data points to assume that feature value.

\begin{figure}[h]
        \vspace{-2mm}
  \centering
  \includegraphics[width=0.8\linewidth]{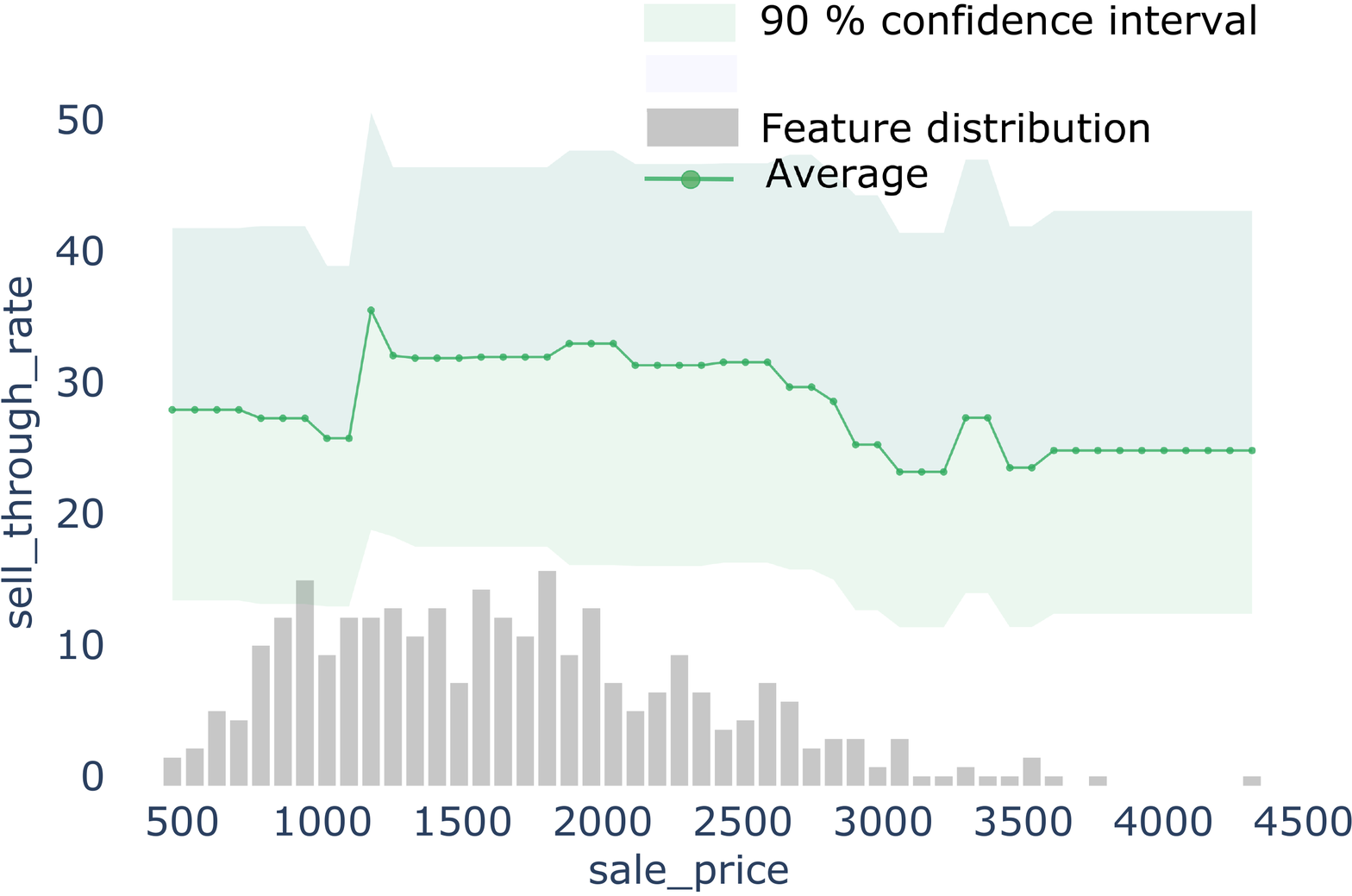}
        \vspace{-4mm}
  \caption{Sample SHAP based partial dependence plot}
  \label{fig:pdpdiscount}
    \vspace{-5mm}
\end{figure}

\noindent \textbf{SHAP based approach}:
If we plot the values of a feature on the x-axis and the corresponding SHAP values of the same feature on the y-axis, then we can generate a partial dependence plot for that feature. This plot shows how the model depends on the given feature, and is like a richer extension of the classical partial dependence plots. Figure \ref{fig:pdpdiscount} shows this plot for {sale\_price}. The calculation for the partial dependence plots has a {\it causal interpretation} too. One way to think about PDP is that it is an {\it intervention query}. We intervene on a feature and measure the changes in the predictions. In doing so, we analyze the causal relationship between the feature and the prediction \cite{zhao2019causal}. The assumption of independence is the biggest issue with PD plots. It is assumed that the feature(s) for which the partial dependence is computed are not correlated with other features.

\vspace{-2mm}
\subsection{Factors affecting sales for a particular product:  Local Explainability} 
Local explainability provides human interpretable explanation on why a particular SKU did/did not sell well. Why did the model make a certain prediction for an instance? We provide designers with multiple tools to answer these questions.

\begin{figure*}[t]
  \centering
  \includegraphics[width=\linewidth]{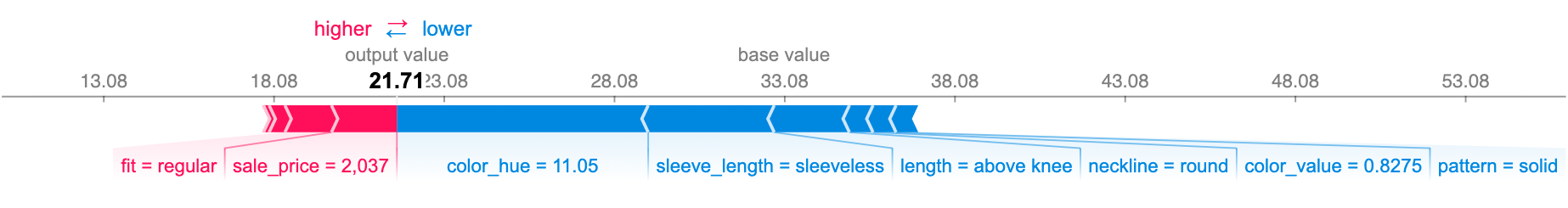}
  \vspace{-5mm}
  \caption{SHAP plot for the sample dress}
  \label{fig:shap}
\vspace{-3mm}
\end{figure*}

\subsubsection{SHAP explanation}
\begin{figure}[h]
  \centering
  \includegraphics[width=0.5\linewidth]{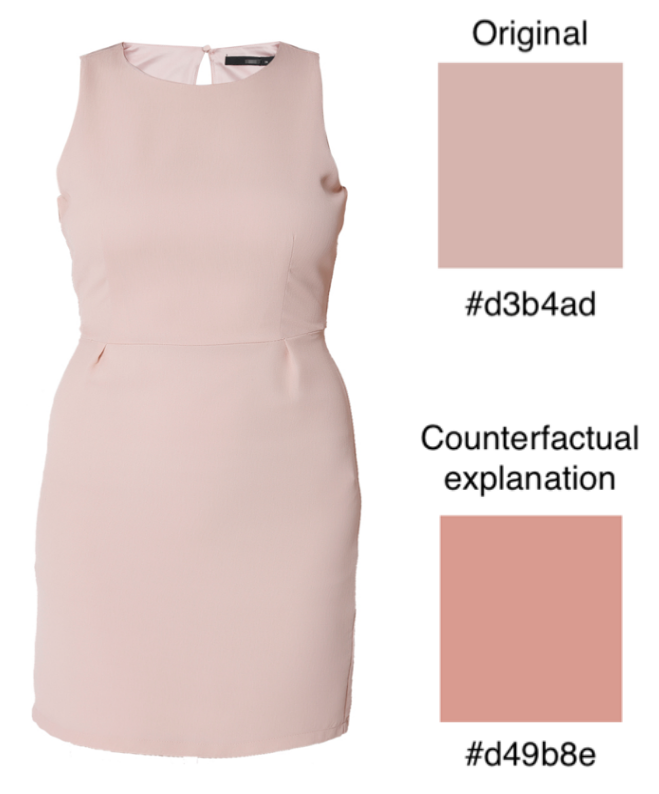}
  \vspace{-5mm}
  \caption{Sample dress for SHAP and counterfactual explanations}
\label{fig:sample_dress}
\vspace{-2mm}
\end{figure}

\begin{figure}[h]
  \vspace{-3mm}
  \centering
  \includegraphics[width=0.5\linewidth]{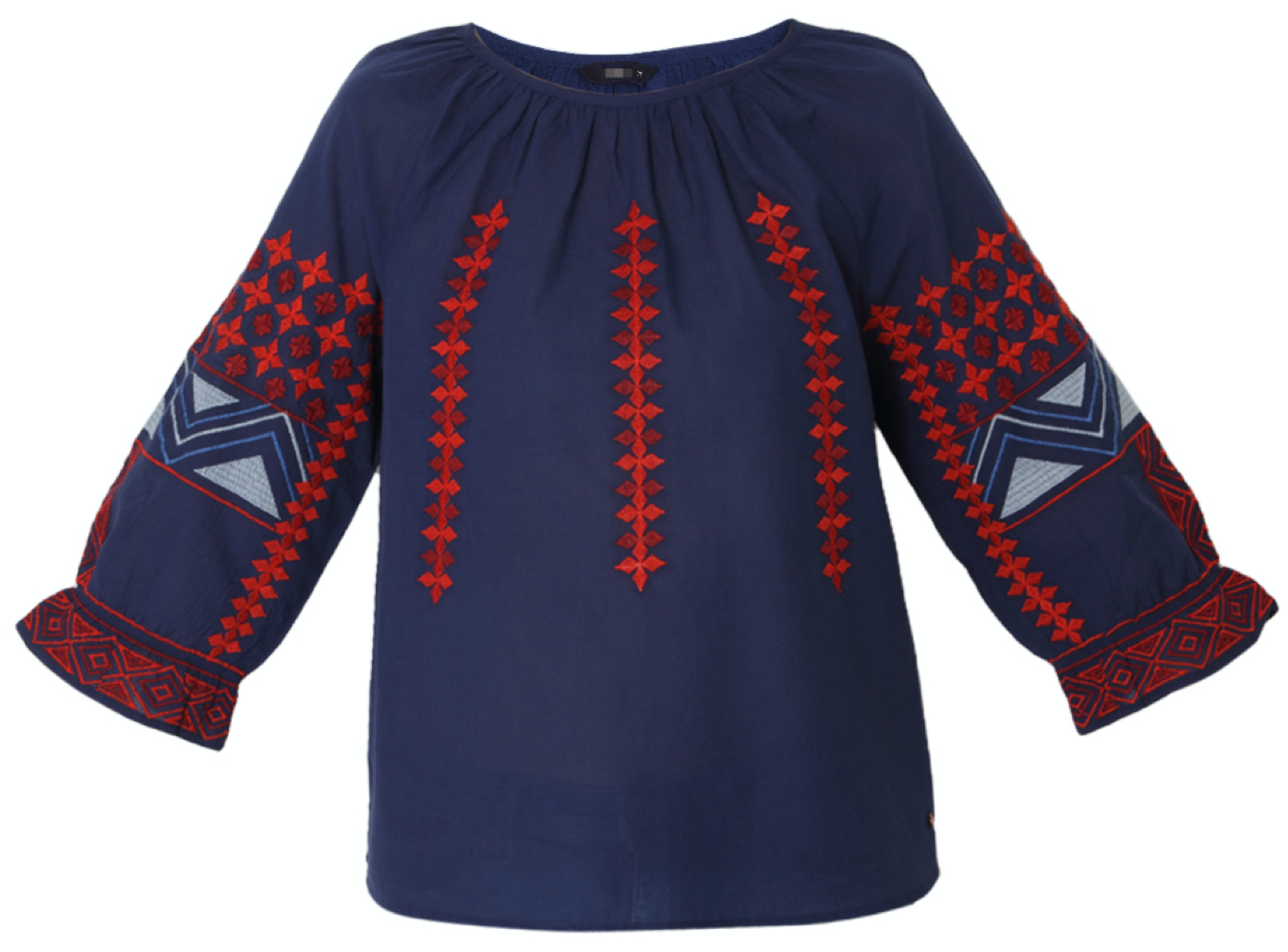}
  \vspace{-3mm}
  \caption{Sample top for SHAP and counterfactual explanations}
\label{fig:sample_top}
  \vspace{-5mm}
\end{figure}

\begin{figure*}[t]
  \centering
  \includegraphics[width=\linewidth]{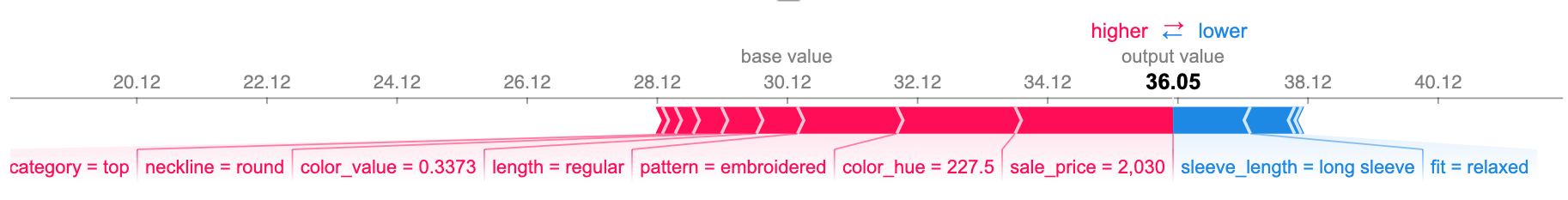}
  \vspace{-7mm}
  \caption{SHAP plot for the sample top}
  \label{fig:shap_top}
\vspace{-2mm}
\end{figure*}

SHAP values were introduced in sub-section \ref{subsection:global} to understand global explainability, however, SHAP values were primarily proposed for local explainability by \cite{lundberg2017unified}. SHAP values can explain the output of any machine learning model but  high-speed exact algorithms are available for tree ensemble methods. In figure \ref{fig:shap}, we show a sample SHAP plot for a product from the {\bf dress} category (shown in figure \ref{fig:sample_dress}). This dress had a STR (21.70\%), 1.37 units lower than the average STR (23.08\%) mainly because of the {color\_hue} (11.05),  {sleeve\_length} (sleeveless) and {length} (above knee) while {sale\_price} (2037) was trying to push it higher. This level of feature attribution helps designers/buyers understand why a product failed in the previous season. For the in season scenario, local explainability for a product aids designers in coming up with better combination of attributes that will have a higher STR. It will further help improve transparency between designers and buyers, because buyers can now rely on local SHAP explanations to explain why a certain design is selected/rejected by them. In figure \ref{fig:sample_top}, we consider the example of sample top which performed better than the average of the catalog and plot its SHAP values in figure \ref{fig:shap_top}.

\subsubsection{Counterfactual query} 
\begin{figure}[h]
  \centering
  \includegraphics[width=0.85\linewidth]{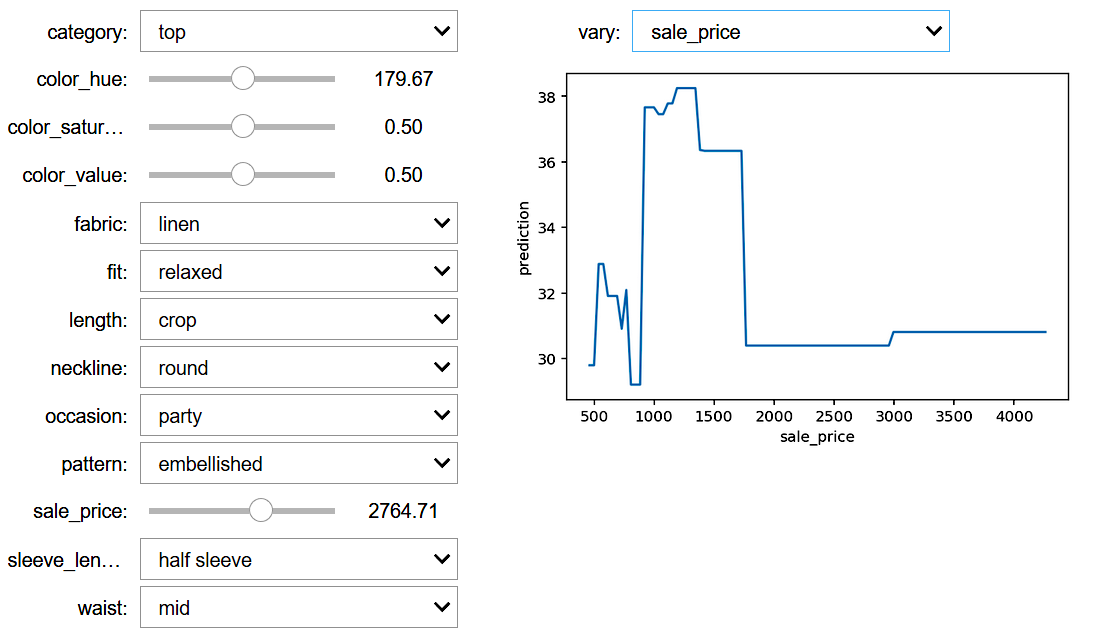}
    \vspace{-3mm}
  \caption{Counterfactual query engine}
  \label{fig:count_query}
    \vspace{-5mm}
\end{figure}

Given a particular product, what is the impact on the target variable of this product if we vary a particular product feature? This is a counterfactual query and designers can use responses to such queries to make important pre-season design decisions. Contrary to partial dependence plots, counterfactual queries are concerned with local explanations, hence they are more useful for individual product design. Figure \ref{fig:count_query} shows a counterfactual query response engine built with drop down buttons for categorical features and sliders for numerical features.

\subsubsection{Counterfactual explanation}
A counterfactual explanation of a prediction describes the smallest change to the feature values that changes the prediction to a predefined output. A counterfactual explanation takes the form of a statement like, “You were denied a loan because your annual income was 30,000. If your income had been 45,000, you would have been offered a loan.”  According to \cite{molnar2019} and \cite{wachter2017counterfactual} given an input $\mathbf{x}$, a model $f$, and a distance metric $d$, the
counterfactual explanation $\mathbf{x}'$ can be obtained by solving the optimization problem:
$$\min_{\mathbf{x}'}\max_{\lambda} \lambda (f(\mathbf{x}')-y')^2 + d(\mathbf{x},\mathbf{x}') $$
where $y'$ is the desired outcome. A higher value of $\lambda$ implies closer proximity to the desired outcome $y'$, whereas a lower value of lambda results in a $\mathbf{x}'$ which is more faithful to the original input $\mathbf{x}$. We explored two model-agnostic, gradient-free approaches to solve this constrained nonlinear optimization problem:
Customized genetic algorithms and Constrained Optimization By Linear Approximation (COBYLA) \cite{powell2007view}. These approaches are not restricted to tree based learners and they do not impose constraints on differentiability of distance functions. Even though COBYLA is a powerful method to solve this problem, it works only with numerical variables, hence a numerical encoder is necessary. However, the optimal solutions obtained using COBYLA may not be invertible back to categorical variables. That is, the optimal solution may not be realistic or practically implementable. Hence, we use our custom genetic algorithms to solve the problem of counterfactual optimization. To illustrate with an example, we consider the dress shown in figure \ref{fig:sample_dress} with a given feature set and a STR of 21.70\%. We set the target STR at 60\% and solve the optimization problem with the resulting solution (counterfactual explanation) shown in table \ref{tab:counter_dress}.
\begin{table}[h!]
  \begin{center}
      \vspace{-3mm}
    \caption{Counterfactual explanation for the sample dress}
    \label{tab:table1}
        \vspace{-3mm}
    \begin{tabular}{l|c c c} 
      \textbf{Product features } & \textbf{Feature} & &\textbf{Counterfactual }\\
      & \textbf{inputs} &   &\textbf{explanation }\\
      \hline

      color\_hue  & 11.0526 & $\longrightarrow$ & 11.330\\
      color\_saturation & 0.1800 &$\longrightarrow$ & 0.2465\\
      color\_value  & 0.8274 &$\longrightarrow$ & 0.8712\\
      length  & above knee &$\longrightarrow$ & regular\\
      neckline  & round &$\longrightarrow$ & strap\\
      sale\_price  & 2036.5 & $\longrightarrow$ & 2036.6\\
            \hline
      \textbf{STR forecast: } & 21.70\% & & 60.56\%
    \end{tabular}\label{tab:counter_dress}
  \end{center}
    \vspace{-4mm}
\end{table}
\noindent The counterfactual explanation results in modification of 5 features: {color\_hue, color\_saturation, color\_value, length, neckline} and {sale\_price}. Since, the change in {sale\_price} is negligible, we neglect it. Modifications for  {color\_hue, color\_saturation, color\_value} results in a new color for the counterfactual explanation given by HEX value of \#d3b4ad. Original color and the color corresponding to  counterfactual explanation are shown in figure \ref{fig:sample_dress}.  
If this optimization problem is solved while constraining some features from getting modified and allowing only the following features to get modified:
 `pattern', `fit', `length', `fabric',  `neckline' (see table \ref{tab:counter_dress2}). We consider a second example (figure \ref{fig:sample_top}) which already performs better than average catalog STR (see SHAP plot in figure \ref{fig:shap_top}). We set the target STR at 60\% and solve the optimization problem resulting in counterfactual explanation shown in table \ref{tab:counter_dress3}.

\vspace{-1mm}
\begin{table}[h!]
  \begin{center}
    \caption{Restricted counterfactual explanation for the sample dress}
    \label{tab:table2}
        \vspace{-4mm}
    \begin{tabular}{l|c c c} 
      \textbf{Product features } & \textbf{Feature} & &\textbf{Counterfactual }\\
      & \textbf{inputs} &   &\textbf{explanation }\\
      \hline

      pattern  & solid & $\longrightarrow$ & striped\\
      length & above knee &$\longrightarrow$ & regular\\
      neckline  & round &$\longrightarrow$ & button down\\
            \hline
      \textbf{STR forecast: } & 21.70\% & & 58.65\%
    \end{tabular}\label{tab:counter_dress2}
  \end{center}
    \vspace{-5mm}
\end{table}

\begin{table}[h!]
  \begin{center}
    \caption{Counterfactual explanation for the sample top}
    \label{tab:table3}
    \vspace{-3mm}
    \begin{tabular}{l|c c c} 
      \textbf{Product features } & \textbf{Feature} & &\textbf{Counterfactual }\\
      & \textbf{inputs} &   &\textbf{explanation }\\
      \hline
      sleeve length & long sleeve &$\longrightarrow$ & sleeveless\\
            \hline
      \textbf{STR forecast: } & 36.05\% & & 42.13\%
    \end{tabular}\label{tab:counter_dress3}
  \end{center}
    \vspace{-5mm}
\end{table}

\noindent It can be observed that the results of counterfactual explanations are in agreement with results from SHAP values. For example color\_hue was responsible for pulling the STR forecast down for the sample dress and the counterfactual explanation suggests a change in color\_hue to improve its STR forecast. Similarly for the sample top, sleeve\_length was the main feature pulling its STR forecast down and the counterfactual explanation suggests a change in sleeve\_length.








\begin{figure}[h]
   \centering
   \includegraphics[width=1\linewidth]{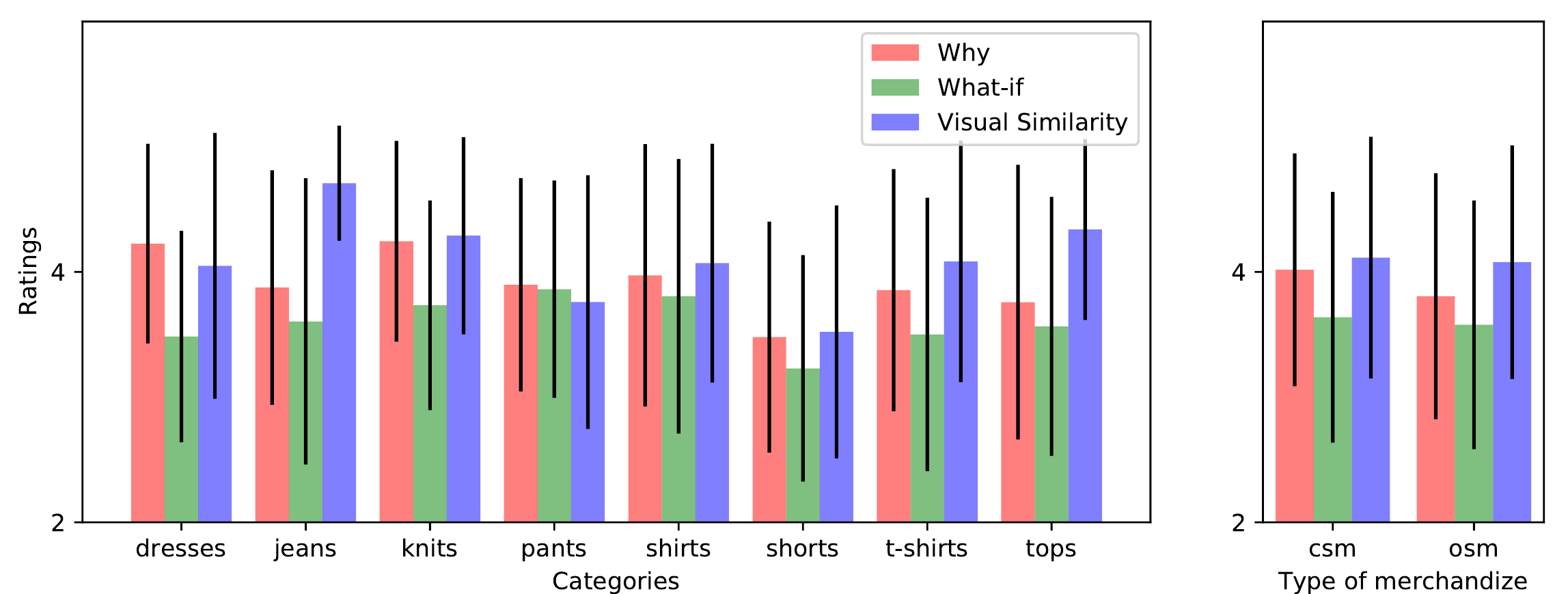}
   \vspace{-5mm}
   \caption{Aggregate ratings from designers and buyers.}
   \label{fig:user_feedback}
   \vspace{-5mm}
 \end{figure}
\vspace{-3mm}
\section{Deployment in industry and persona specific use cases} \label{sec:use_cases}
The explainable sales forecasting system has been deployed for our industry partner and is currently under use for past-season analysis and pre-season planning for the next Autumn Winter (AW) season. Even though improving sell through rates and reducing dead inventory are the final goals of our system, our initial focus has been adoption of our AI tools by designers and buyers. Our system makes sure that increased adoption of our AI tools by designers in the pre-season phase will enable the designers to design better selling products (based on new product forecasts) and also generate high { hit-rates}. High hit-rates are indicative of improved buyer designer collaboration and it is an important criteria based upon  which the designers are evaluated by the {end of season}.  Also, our system enables buyers to develop optimal product range plan which caters to the needs of initial option plan designed by planners. The current User Interface (UI) for our system has been designed to address the needs of two stakeholders: designers and buyers. Hence, we have a designer view and buyer view in the UI. 

\noindent {\bf Designer view}: Based on our consultations with designers, a {\it product life cycle} view for all the various steps in the designer's work flow have been provided. Product life cycle view starts with a designer landing page which enables designers to upload inspirational images, design images and product attributes of a new design. Based on an initial new product forecast and a screening by management, the new design idea moves to the next stage, where the designer is required to enter more detailed attributes of the new design. This stage is accompanied with the images and details of all the visually similar products (w.r.t. new design) from past season catalog.  Apart from a forecast for the newly designed product, the designer is further enabled with the following capabilities:
\begin{itemize}
\item Explainable new product sales forecast (global and local).
\item PDP plots and counterfactual query plots of product attributes for what-if analysis and counterfactual explanations.
\item Explainable sales analysis for visually similar products from past seasons.
\end{itemize}
This stage also carries a like button and feedback form for each design. The feedback/comment form is used by all the stakeholders to provide critical feedback and pursue conversations regarding the new design. This stage also consists of a manual assortment selection tool for the buyers. Buyers use this tool to express their interest in acquiring the new product under design. An important challenge faced during planning and deployment of our system was the manual effort required by designers to upload detailed attribute data. This problem was addressed by incentivising designers with a product life cycle view (with AI insights) and capabilities to export designs into spreadsheets, slides and portable document format.
\noindent{\bf Buyer and merchandiser view}: This view consists of a different landing page whereby buyers are provided global view of all the products and their performance at different levels of granularity for both current season and past seasons. They are enabled to search the product catalogs by SKU/category/free text/store id. Search by free text is enabled by natural language semantic search \cite{uren2007usability}. The search results are in the form of 
\begin{itemize}
\item histograms (mean, median) of overall sales/sell-through-rate,
\item actual sales, stock, price time-series,
\item best and worst sellers product list.
\end{itemize}
\noindent We also provide drop down filters like category/attributes, seasons and region to improve search results. Clicking on a particular product from the best and worst sellers product list will land the buyer/user on the {\bf product view}. It consists of 
\begin{itemize}
    \item product images, product design and merchandising attributes,
    \item actual total sales, stock, sell-through-rate, actual sales, stock, price time-series.
\end{itemize} 
\noindent Product view consists of following insights for a product selected from current season:
\begin{itemize}
 \item sales forecast, sell-through-rate for the next $x$ weeks,
\item explanation of the current sales and forecast (local explainability for each SKU),
\item visual Browse to show other similar products from the current season and past seasons. 
\item sales/sell-through-rate across different stores on a map view.
\end{itemize} 
\noindent Product view will be enabled with optimal markdown interventions and markdown simulator for future work. Clicking on a store in the map view lands the buyer/user on the {\bf Store view}. It is enabled with all the features of buyer and merchandiser view (with all the products restricted to the corresponding store). Similarly {\bf Product $\times$ Store view} is a restricted version of the product view with data restricted only to that store.

\subsection{User feedback analysis}
We conducted a limited feedback survey with our industry partner to test how rubber meets the road. We obtained both qualitative and quantitative feedback from users. For quantitative feedback, we randomly selected  3 designers and  3 buyers from the firm and asked them to rate our system's performance for 100 randomly selected products from their current and old season merchandise. The survey was carried out for 8 categories of products  for their sales at the national level and the system was rated on its three core AI modules: 
\begin{itemize}
\item SHAP Explanations  for product sales (\texttt{Why}).
\item Counterfactual query plots of product attributes for what-if analysis (\texttt{What-if}).
\item Visually similar products from past seasons (\texttt{Visual Simiarity}).
\end{itemize}
The rating was on a scale of 1 to 5, with 1 representing complete disagreement  and 5 representing complete agreement with explanations. For \texttt{Visual Simiarity}, we obtained ratings based its perceived usefulness. The results of the survey are summarized in figure \ref{fig:user_feedback}.  The plot reveals that users (designers and buyers) found \texttt{Visual Simiarity} to be generally useful. With respect to  explanations, their trust in \texttt{Why} seems to consistently higher than \texttt{What-if}.  It can also be observed that ratings for visual similarity vary significantly with category with highest results for jeans. However, nature of merchandise: current or old season, does not have significant on the ratings for all 3 modules.\newline

\noindent Apart from gathering ratings, thorough discussions were held with users for qualitative feedback. 

\noindent \textbf{Designer feedback}:
Designers saw this tool as validation mechanism for their gut instinct about a product. But they were mainly appreciative of the improved workflow provided by our AI infused product life cycle view which integrated multiple databases, excel files, power point presentations, portable document format files into single location with search capabilities. However, designers were concerned about non-inclusion of influencing variables like 
\begin{itemize}
    \item store-level visual merchandising including rack arrangement and planograms,
    \item visually similar products in competitor catalogues. 
\end{itemize}
 Another important feedback from designers involved request for a tool that analyses how anchoring few features influences selection of other features. This form of explainability has been addressed by \cite{ribeiro2018anchors} by  formulating it as a pure exploration multi-armed bandit problem and it will be considered for inclusion in future versions of this system.

\noindent \textbf{Buyer feedback}: Buyers championed this tool right from its inception. This tool definitely made product sales review easier and allowed buyers to analyse historical sales trends and new product forecasts quickly without any ambiguity in one location. This tool primarily provided them a good head start for a new range plan. Buyers were also appreciative of the visually similar products functionality, which showed how visually similar product performed in the past seasons. Though the system provides a new product forecast, buyers could confirm these predictions using the past performance of visually similar products.





\vspace{-3mm}
\section{Conclusions and future work}\label{sec:conclusion}
The topics of new product forecasting and explainability have received  increased attention over recent years, however, there aren't many works on incorporating explainability and interventions for new product forecasting in the context of sustainable fashion. In this paper, we presented a detailed case study on designing and deploying an explainable AI based new product forecasting system for fashion industry.  Apart from explainability, important problems of data integration, following data hygiene and providing data export capabilities were solved during the project and these aspects also improved the adoption of our system by the stakeholders. In this paper, we restrict ourselves to pre-season planning and interventions, which involves product development and design by designers and preparation of final product range by buyers. However, efficient pre-season planning is the first step towards sustainable fashion, because it enables fashion houses to optimize product design, development and sourcing such that the chances of unsold dead inventory are reduced. We are currently exploring the use of counterfactual explanations to improve sustainability indices of products. We are also working on other pre-season aspects like hyper-local assortment generation and stock allocation as a part of our future work.


\vspace{-1mm}
\bibliographystyle{ACM-Reference-Format}
\bibliography{sample-base}

\end{document}